# The effects of convection criteria on the evolution of population III stars and the detectability of their supernovae

T.M. Lawlor[1], T.R. Young[2], J. Teffs[2] and J. MacDonald[3]

[1]Pennsylvania State University, Brandywine, Department of Physics, Media, PA 19063 USA
[2]University of North Dakota, Department of Physics, Grand Forks, ND, 58202, USA
[3]University of Delaware, Department of Physics and Astronomy, Newark, DE 19716, USA

**ABSTRACT**

The first stars continue to elude modern telescopes, but much has been accomplished in observing the glow of the first galaxies. As detection capabilities improve we will eventually resolve these galaxies, but hopes of observing an individual star remains dim for the foreseeable future. However, our first view of an individual first star may be possible due to its explosion. In this work, we present evolution calculations for Population III (Pop III) stars and their subsequent supernovae explosions. Our evolution models include a mass range of 15 - 100 $M_\odot$, each with initial heavy element abundance $Z = 10^{-14}$. Our models are evolved from pre-main sequence through formation of an iron core, and thus near to core collapse. We find that modelling the evolution of these stars is very sensitive to the choice of convection criterion; here we provide evolution results using both the Schwarzschild and Ledoux criteria. We also use the final structure from our models for numerical simulation of their supernovae light curves using a radiation hydrodynamics code. In doing so, we estimate a lower bound of initial model mass that may be possible to observe in the near future. We also find that our 40 $M_\odot$ Schwarzschild evolution model produces the brightest supernova peak and statistically should be the most frequently observed. At our highest redshift $z= 15$, only the 60 $M_\odot$ Schwarzschild model at peak magnitude starts to rival the 40 $M_\odot$ model in brightness.

## 1 INTRODUCTION

One important reason to study first stars is that they are the likely cause of re-ionization of hydrogen in the early universe. The epoch known as re-ionization lasted for a time between redshift $z = 10$ and 6. This is a time span of 500 – 700 Myr after the Big Bang, or about 12.8 – 13.5 billion ly away (Tanvir et al., 2009). Currently, the furthest found galaxy is at about redshift $z = 11$ (Coe et al. 2013). More galaxies have been found in the range $z = 7 - 8$, which corresponds to a time of about 800 Myr after the Big Bang. It has been suggested that the first stars began forming as early as 200 Myr after the Big Bang (Glover 2005; Bromm & Larson 2004) at $z \approx 20$, thus we may have not yet looked far enough away to see Pop III stars, although more recently it was reported that formation may not have begun until as a late as 100 Myr later (Plank Collaboration, 2014). The early universe stars are expected to have formed in a very different environment than what we see today, one that was primarily hydrogen and helium. It has been reported until fairly recently that the first stars were exclusively very massive single stars (Heger & Woosley 2002). While it is still expected that the first stars should have been quite massive, some recent modelling results have shown that lower mass binary systems may have formed (Turk, Abel & O'Shea 2009; Stacy et al. 2010). It was proposed early on (Palla, Salpeter & Stahler 1983; Eryurt 1985) that there should be a wide range of masses for Pop III stars, with a distribution that is bimodal around 1 $M_\odot$ and 100 $M_\odot$. Cai et al. (2011) found by measuring He II $\lambda$1640 emission in the distant galaxy IOK-1 at $z = 6.96$, that massive Pop III stars represent less than 6% of the total star formation, though they note many uncertainties such as an unknown IMF and uncertainty in the details of Pop III stellar evolution.

No Pop III stars have been conclusively identified, possibly because stars of low enough mass to survive to the present day do not form (Bromm & Loeb 2003; Schneider et al. 2003) or else are polluted during formation or subsequent evolution. Because of their short lifetimes, bright massive Pop III stars reside at high redshift $z > 10$. There are a number of investigations underway or planned to observe objects that may indirectly shed light on these early universe objects. Among instruments available for this purpose include the Hubble Space Telescope (HST) /Wide Filed Camera 3 (WFC3)*.* Tens of galaxies have been imaged between redshift $z = 7$ and 8 and are described by Bouwens et al. (2010), for example. Oesch et al. (2010) report that some of these galaxies are irregularly shaped and small compared to today's counterparts. Other instruments being used to this end include the Spitzer Space Telescope, the Cosmic Infrared Background Experiment (CIBER) and the Swift Observatory. Swift is used to discover and observe gamma ray bursts, events that are very energetic and thus possible to see in very distant galaxies. Two observing projects that will push the limits of distance are the Atacama Large Millimetre/submillimetre Array (ALMA, operational as of 2013) and of course the James Webb Space Telescope (JWST, projected to be launched in 2018). The JWST is expected to see as far as redshift 15 without gravitational lensing (Rydberg et al. 2013; Zackrisson et al. 2012; Dunlop 2012; Whalen et al. 2012). Rydberg et al. (2013) suggest that while the JWST may be able to detect Pop III galaxies, the hope of seeing Pop III stars individually remains dim. They find that in a non-lensed field a 60 $M_\odot$ star would have a magnitude that is six magnitudes below the detection capabilities of JWST. Detection does become just possible through a gravitational lens, but with a rather high gravitational magnification needed. The best possible case for detection is for a very massive star (about 300 $M_\odot$) with a very high gravitational magnification, but the authors concede that even this ideal case is not very likely.

There is a growing body of theoretical evolution calculations for Pop III stars. The evolution of Pop III stars, including the effects of binary interaction is described by Lawlor et al. (2008) using the evolution code BRAHMA (Lawlor & MacDonald 2006). Model stars were evolved from the pre-main sequence through the tip of the giant branch, and to their resulting supernovae explosions. The code used for supernovae simulations is a one-dimensional Lagrangian radiation hydrodynamics code (Sutherland & Wheeler 1984; Young 2004). This work uncovered a number of differences between binary and single star evolution. It further showed a trend that Pop III supernovae show a fainter peak and longer plateau than their later metal rich counterparts. Limongi & Chieffi (2012) present a grid of evolution models using the FRANEC code (Limongi & Chieffi 2006), and calculations of nucleosynthetic yields during explosions for model masses between 13 $M_\odot$ and 80 $M_\odot$. Earlier, Chieffi et al. (2003) evolved models between 13 $M_\odot$ and 25 $M_\odot$ for a range of $Z$ at constant mass. They also calculate light curves for their grid of models and found similar results as those of Lawlor et al. (2008) in that light curves for very low metal models exhibit a dimmer light curve. This is generally due to lower metal stars having smaller radii and light curves depend sensitively on the initial stellar radius (Young 2004). A similar analysis for evolution and light curves of pair instability supernovae (PISN) for models of masses 150 $M_\odot$ and 250 $M_\odot$ is provided by Whalen et al. (2012). Eckström et al. (2010) followed the evolution of 15 $M_\odot$ – 60 $M_\odot$ models up to the end of helium core burning. They specifically focus on the effects of varying fundamental constants. Finally, the evolution of zero metallicity stars at constant mass for a range up to 100 $M_\odot$ is described by Marigo et al. (2001).

We focus here on the premise that if a single star is six magnitudes below the JWST's detection limit then it may be possible to observe the much brighter supernova explosion of an individual first star. We test the sensitivity of detectability estimates to one aspect of modelling the prior evolution, namely the treatment of convective energy transfer and mixing. In section 2, we present stellar evolution calculations for masses between 15 $M_\odot$ - 100 $M_\odot$ for Pop III stars from pre-main sequence to very near core collapse. Two models for convection are used; one based on the Schwarzschild criterion for convective onset, and the other on the Ledoux criterion for convective onset. In section 3, we use the final stellar structure for a selection our evolution models to calculate resulting supernovae light curves using a 1-D radiation hydrodynamics code. We present light curves both un-attenuated and attenuated using red shift and k-corrections as described by Kim et al. (1996) to account for dimming due to large cosmological redshifts of $z = 15$.

## 2 EVOLUTION TO THE IRON CORE

We have evolved non-rotating Pop. III stellar models from the pre-main sequence to the stage at which photo-destruction of iron-peak elements in the core begins. We assume that shortly after this point, the star will core collapse and proceed to explode. The stellar masses are in the range 15 – 100 $M_\odot$. The upper limit of the range is selected to avoid pair instability. Our stellar evolution code (BRAHMA) used for model evolution calculations is described in detail by Lawlor & MacDonald (2006) and updates pertinent to stellar evolution for high mass stars are described by Lawlor et al. (2008). To determine how the treatment of convection affects the evolution, we have calculated models using the Ledoux and the Schwarzschild criteria for convection. We do not include semi-convection or convective overshoot.

Because of the very low primordial heavy element abundances mass loss through radiative driven mass loss is negligible in our models of Pop III stars. However, some of our models become cool enough that mass loss from winds driven by heating from waves generated in the outer convective or by radiation pressure on grains may occur. We have included mass loss using a Reimers' (1975) formula appropriate to RGB stars together with a fit to the mass loss rates of Mira variables and OH/IR sources appropriate to stars that have experienced dredge-up of heavy elements (Lawlor & MacDonald 2006). If dredge-up occurs, it does so after the end of core He burning, which leaves a relatively short time, of about $10^4$ yr, for mass loss to have any effect on the stellar mass. We find that the mass loss rates never exceed $10^{-5}$ $M_\odot$ yr$^{-1}$, and conclude that these stars would return only modest amounts of heavy elements to the ISM before they explode as supernovae.

## 2.1 Convection treatments

To determine the convective energy flux in models based on the Schwarzschild criterion for convective onset,

$$\nabla > \nabla_{ad}, \qquad (1)$$

we use standard mixing length theory (Böhm-Vitense 1958) with a modification to include radiative losses from convective elements when they are optically thin (Mihalas 1978). Convective mixing is modelled as a diffusion process, with the diffusion coefficient determined from the mixing length theory. For models based on the Ledoux criterion in which a molecular weight gradient has a stabilizing effect,

$$\nabla > \nabla_L \equiv \nabla_{ad} + \frac{\beta}{4-3\beta}\frac{d\ln\mu}{d\ln P}, \qquad (2)$$

we use the same mixing length theory as the Schwarzschild models except that $\nabla_{ad}$ is replaced everywhere by $\nabla_L$.

A local linear analysis by Kato (1966) showed that regions of the star in which $\nabla_{ad} < \nabla < \nabla_L$ are secularly overstable. The growth time of the instability, $\tau_{so}$, is of order the thermal time scale of the convective element. If the stellar evolution time scale is short/long compared to $\tau_{so}$, then the temperature gradient from the Ledoux/Schwarzschild convection model is appropriate. If the stellar evolution time scale is of order $\tau_{so}$, then we expect the temperature gradient to be intermediate to the two limiting cases. Secularly overstable regions of a star are often referred to as being semi-convective, and we will use that terminology here. A number of recipes for treating mixing in semi-convective regions have been developed (e.g. Weaver, Zimmerman &

Woosley 1978; Langer, Sugimoto & Fricke 1983; Maeder 1997). In addition to mixing, Spruit (1992) has also included consideration of the energy flux in semi-convective regions. Rather than implement a particular recipe for semi-convection, we consider the two limiting cases in which $\tau_{so}$ is assumed either much smaller or much larger than the stellar evolution time.

The term semi-convection is also used in Schwarzschild convection when convective mixing, through its effect on opacity, leads to a composition profile such that $\nabla = \nabla_{rad} = \nabla_{ad}$ (Eggleton 1972). To differentiate from the above use of semi-convection, we will refer to such regions of a star as being marginally convective.

## 2.2 Stellar evolution results

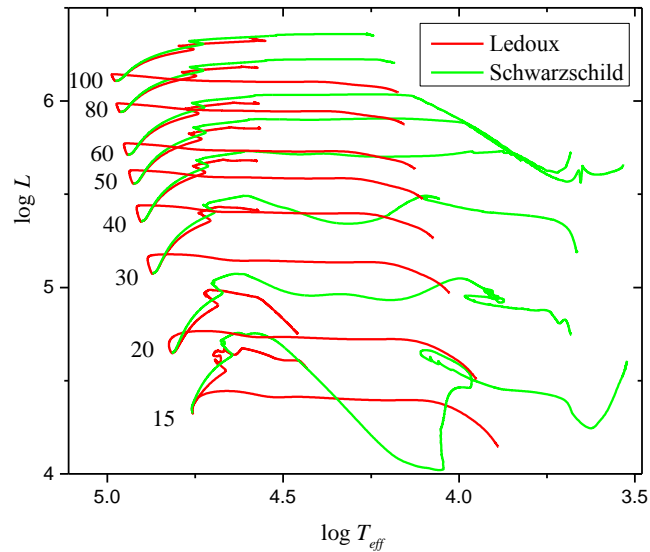

**Figure 1.** Evolutionary tracks in the HR diagram. For clarity the pre-main sequence phase for the Schwarzschild convection models is not shown. Labels shown are in solar masses.

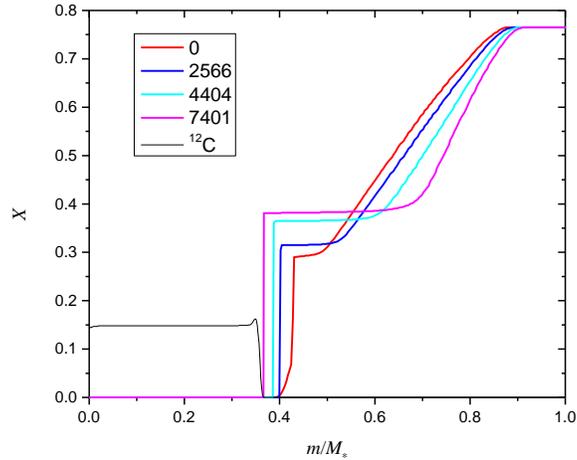

**Figure 2.** Evolution of the H profile in a 60 $M_\odot$ Schwarzschild convection model. The legend shows elapsed time in years. Also shown is the $^{12}$C profile.

Since the early evolution of pop. III stars has been extensively studied (Ezer & Cameron 1971; Marigo et al. 2001; Siess, Livio & Lattanzio 2002), we mainly focus on the evolution after core helium burning. It is at this point that the effects of different treatments of convection are most marked. From figure 1, which shows evolutionary tracks in the HR diagram, it is clear that the Schwarzschild convection models reach much larger radii and much lower effective temperatures at core collapse than their Ledoux convection counterparts. At the end of core helium burning, there is a marginally convective region extending outwards from above the H-burning shell. The hydrogen profile at this point is shown for the 60 $M_\odot$ Schwarzschild convection model by the red line in figure 2. A similar profile is found at this stage for other masses and for the Ledoux convection models. For the Schwarzschild convection models, convective mixing is uninhibited by the molecular weight gradient and the hydrogen profile evolves as shown by the other lines in figure 2. The base of the convection zone moves inwards through the helium layer until it reaches carbon and oxygen-rich regions. At this point energy generation in the H-burning shell increases dramatically and powers the expansion of the envelope to large radii. In some cases, the surface convection zone reaches the photosphere and dredges heavy elements to the surface. For the Ledoux convection models, the molecular weight gradient hinders development of the convection zone and prevents its base moving inwards to carbon-rich layers. The power from H-burning does not increase sufficiently to cause envelope expansion. To show the differences in behavior of the convection zones, we plot the convection zone boundaries as a function of time in figure 3 for our 60 $M_\odot$ models. Because the later stages of evolution are much faster than the earlier stages, the time coordinate is taken to be the logarithm of the time until the end of the calculation near the point of core collapse.

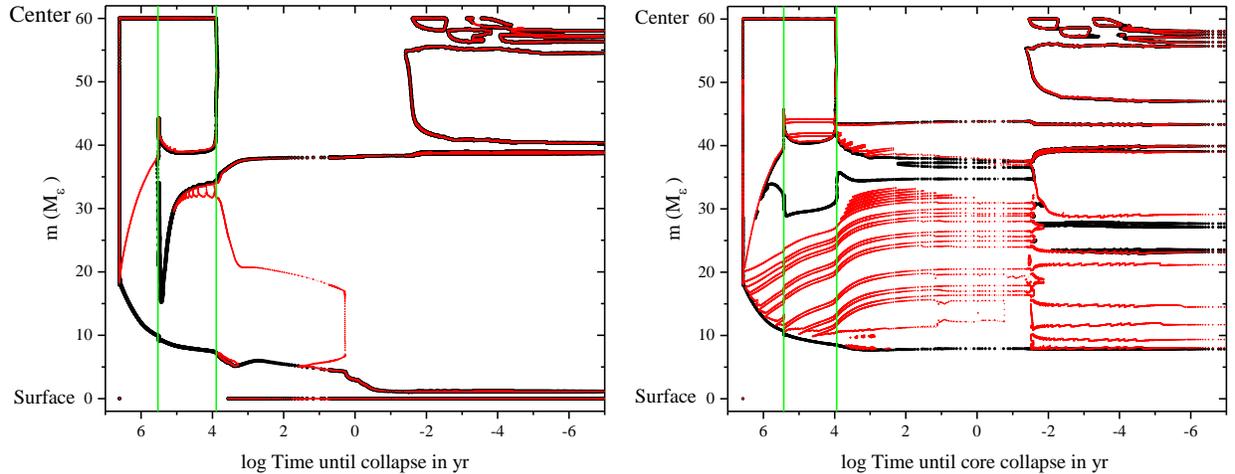

**Figure 3**. Evolution of the convection zone boundaries for our 60 M$_\odot$ Schwarzschild convection model (left panel) and our 60 M$_\odot$ Ledoux convection model (right panel). The black and red dots delineate the convection zone boundaries determined by applying the Schwarzschild and Ledoux criteria for convection onset, respectively.

The left and right panels show the locations of the convection zone boundaries for the evolution using the Schwarzschild and Ledoux convection criteria, respectively. The black and red dots show the boundaries given by the Schwarzschild and Ledoux criteria, respectively. The vertical green lines show the end points of the H and He core burning phases. From the left panel, we see that the core convection zone mixes the products of He burning out as far as 21.25 M$_\odot$, measured from the stellar center. After the end of core He burning, the bottom of the surface convection zone moves inwards until it makes contact with the $^{12}$C-rich region created during core He burning (see fig. 2). By comparing the locations of the Schwarzschild and Ledoux convection zone boundaries, it can be seen that, before and after the end of core He burning, there is a semi-convective region of the star. Even though the Schwarzschild criterion is used for this model, this semi-convective region does not become fully mixed. Instead convective mixing leads to a marginally convective state, as defined in section 2.1. The right panel shows the intricate convection zone structure that results when the Ledoux criterion for convective onset is used. During core H and He burning, the outer parts of the star contain a 'stair-case' of alternating fully-mixed convective regions and radiative regions which are stabilized against convection by a molecular weight gradient. The stair case structure persists after the end of core He burning and becomes more complex. Figure 4 shows the H abundance profile at 3 times: 1) When the star is at the end of the core H burning stage, 2) When the star is at the end of the core He burning stage, and 3) $10^3$ yr before the star's core collapses. Also shown is the C abundance profile at time $10^3$ yr before core collapse.

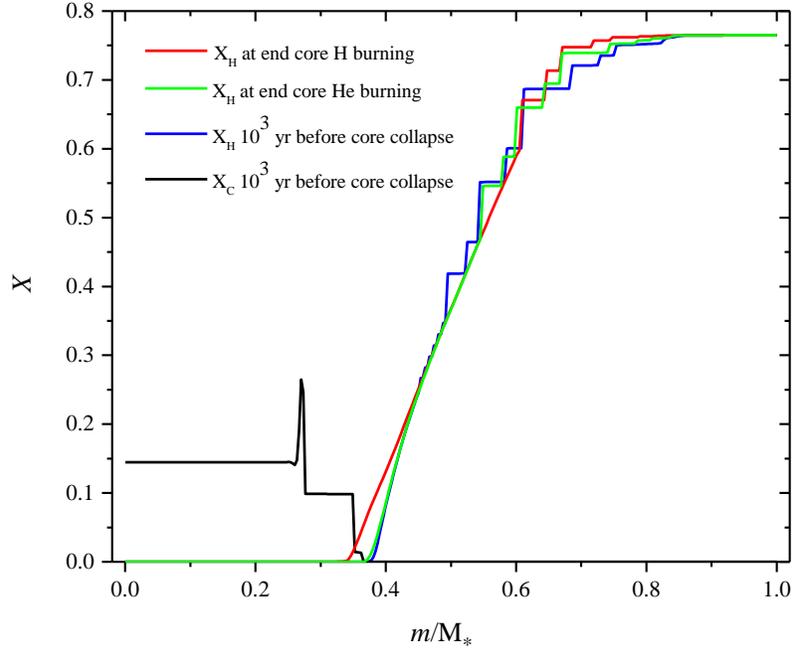

**Figure 4**. Evolution of the H profile in a 60 $M_\odot$ Ledoux convection model. Also shown is the $^{12}$C profile at time $10^3$ yr before core collapse.

Although convection mixes hydrogen inwards in the outer 50% of the star's mass, because of the stabilizing molecular weight gradient, convection does not develop further in and there is no mixing with the carbon containing layers.

In table 1, we compare our results for radius at core-collapse with those found by Heger & Woosley (2010, hereafter HW). HW use Ledoux convection with semi-convective mixing treated as a diffusion process with a diffusion coefficient equal to 10% of the radiative diffusion coefficient, $D_r = 4acT^3/(3\kappa\rho^2 C_P)$. They also include a small amount of convective overshooting by forcing convective boundary zones to be semi-convective. Based on the criteria for convective mixing and its effect on the H-burning luminosity at the end of core He-burning, we expect the radii found by HW to lie between those from our Ledoux and Schwarzschild models. As expected, our Ledoux models all have radii smaller than the HW models. However, for the 50 and 80 $M_\odot$ models, the HW radii are larger than those of our Schwarzschild models. For the more massive HW models that become red supergiants, the processes that lead to expansion to large size are initiated during He burning by the convective core breaching a small entropy barrier and mixing with the surrounding hydrogen containing layers. In our Ledoux convection models of similar mass, the entropy barrier at this stage does become thin but is never breached.

**Table 1.** Radii of pre-supernova models

| Mass/$M_\odot$ | $R_{Ledoux}/R_\odot$ | $R_{Schwarzschild}/R_\odot$ | $R_{HW}/R_\odot$ |
|---|---|---|---|
| 15 | 8.74 | 603 | 10 |
| 20 | 9.57 | 341 | 13 |
| 25 | 12.12 | 567 | 19 |
| 30 | 12.06 | 1066 | 20 |
| 40 | 16.08 | 2048 | 23 |
| 50 | 20.51 | 520 | 2020 |
| 60 | 22.59 | 1038 | 150 |
| 70 | 64.60 | 399 | 184 |
| 80 | 28.09 | 179 | 2334 |
| 100 | 31.57 | 150 | 1.3 |

## 2.3 Convective dredge-up of heavy elements

Convective dredge-up of heavy elements to the surface occurs in our 15, 25, 30, and 40 $M_\odot$ Schwarzschild convection models but not in the Ledoux or our other Schwarzschild convection models. Dredge-up occurs when $T_{eff}$ becomes low enough that the base of the surface convection zone moves inwards to where it meets the convection zone above the H-Burning shell. For the 40 $M_\odot$ model, dredge-up begins before the start of core carbon burning, whereas for the lower mass models dredge-up occurs during core carbon burning. The surface abundances in our pre-supernova models are given in table 2. In each case, oxygen is the major element and, with the exception of the 15 $M_\odot$ model, the mass fraction of N is greater than C. The O survives because the convective turnover time scale for the pressure scale height above the H-burning shell, of order $5\ 10^4$ s, is much shorter than the life time of $^{16}O$ to proton capture. There is a trend of Ne and Mg increasing with stellar mass. The last column of table 2 gives the amount of mass lost due to stellar winds.

**Table 2.** Surface abundances of Schwarzschild convection pre-supernova models in which dredge-up occurs.

| Mass/$M_\odot$ | H | He | C | N | O | Ne | Mg | $\Delta M$ |
|---|---|---|---|---|---|---|---|---|
| 15 | 0.591 | 0.378 | $1.17\ 10^{-2}$ | $4.47\ 10^{-3}$ | $1.46\ 10^{-2}$ | $7.66\ 10^{-7}$ | $2.99\ 10^{-8}$ | $8.1\ 10^{-4}$ |
| 25 | 0.762 | 0.237 | $4.96\ 10^{-5}$ | $6.06\ 10^{-5}$ | $2.46\ 10^{-4}$ | $2.52\ 10^{-7}$ | $1.61\ 10^{-9}$ | $3.2\ 10^{-3}$ |
| 30 | 0.715 | 0.281 | $4.51\ 10^{-4}$ | $5.99\ 10^{-4}$ | $2.56\ 10^{-3}$ | $4.63\ 10^{-6}$ | $3.72\ 10^{-8}$ | $6.3\ 10^{-3}$ |
| 40 | 0.551 | 0.434 | $1.24\ 10^{-3}$ | $2.45\ 10^{-3}$ | $1.05\ 10^{-2}$ | $3.03\ 10^{-5}$ | $3.57\ 10^{-7}$ | $7.1\ 10^{-2}$ |

## 2.4 Pre-supernova models

**Table 3.** Masses for helium and iron cores for models using both the Schwarzschild and Ledoux criteria compared to values from HW.

| Mass/M$_\odot$ | He core Mass/M$_\odot$ | | | Fe core Mass/M$_\odot$ | | |
| --- | --- | --- | --- | --- | --- | --- |
| | Schwarzschild | Ledoux | HW | Schwarzschild | Ledoux | HW |
| 15 | 1.86 | 2.52 | 3.70 | 1.26 | 1.33 | 1.28 |
| 20 | 3.15 | 3.20 | 5.58 | 1.42 | 1.41 | 1.46 |
| 25 | 4.70 | 5.00 | 7.62 | 1.60 | 1.54 | 1.59 |
| 30 | 7.00 | 9.10 | 9.95 | 1.61 | 1.57 | 1.50 |
| 40 | 11.59 | 13.65 | 15.29 | 1.89 | 1.51 | 1.88 |
| 50 | 16.80 | 17.78 | 17.78 | 1.83 | 1.82 | 1.82 |
| 60 | 21.00 | 22.58 | 23.90 | 2.02 | 1.97 | 1.91 |
| 70 | 23.79 | 24.09 | 28.78 | 1.87 | 1.99 | 1.96 |
| 80 | 31.71 | 32.14 | 31.39 | ---$^a$ | 2.16 | 2.14 |

$^a$ Becomes unstable due to pair formation during Si burning

In Table 3 we compare our core masses in the pre-supernova models with those of HW. For the lower mass models, we see that the smallest He core masses are found for our Schwarzschild convection models. This is a consequence of the base of the convection zone moving inwards after core helium burning. Also our Ledoux convection models give He core masses intermediate between those of our Schwarzschild convection models and the HW models. We attribute this behaviour to inclusion of convective overshoot in the models of HW which leads to larger core masses than our Ledoux models which do not include convective overshoot.

With few exceptions, the Fe core masses are found to be insensitive to the treatment of convection. Thus differences in supernova light curves are likely to result from differences in envelope structure rather than the details of the core collapse.

## 3 SUPERNOVA SIMULATIONS

Stellar evolution models are evolved until a significant Fe core is reached and their final structures used as input for our radiation hydrodynamics code to perform supernova calculations. We calculate all physical parameters associated with explosion as well as the observable luminosity versus time. The code, described briefly by Lawlor et al. (2008) and in detail by Young (2004), is a one dimensional Lagrangian radiation hydrodynamics code which is a modified version of the code used in Sutherland & Wheeler (1984). The explosion parameters for each model are followed for 400 days when the luminosity has had sufficient time to reach the constant spontaneous release of energy from $^{56}$Ni decay. The actual collapse of the Fe/Ni core is not followed in this work. For simplicity we remove a gravitational mass of 1.6 $M_\odot$ for all models and replace it with a gravitational point mass. That mass is removed and set as an inner boundary condition in the light curve models. The canonical total gravitational potential energy for this mass is $1 \cdot 10^{53}$ ergs. Explosions are simulated by placing an equivalent energy at the inner most zone of the model, producing explosions at kinetic energies between $1 - 5 \cdot 10^{51}$ ergs, although the more massive models need more energy to produce explosions. For example, the 80 $M_\odot$ models need $8 \cdot 10^{51}$ ergs. The main shock wave is followed through the star producing smaller reverse shocks at compositional boundaries. The forward shock leaves the Lagrangian grid, setting the model into an expansion. After about 100 days, these ejecta settle into a homologous expansion persistent to 400 days.

In figure 5 we show as an example, two chemical composition profiles for both convection criteria, discussed in section 2, for the 15 $M_\odot$ and 70 $M_\odot$ models, both at the end of their evolution. Ledoux convection models are shown in the top panel and the Schwarzschild convection models in the bottom. The neutron star or black hole mass is accounted for in the plot as a blank space up to 1.6 $M_\odot$, and the composition lines begin at the edge of the compact object. For model masses in the range $15 - 40$ $M_\odot$, the Schwarzschild models have less hydrogen in the envelope than the Ledoux models, but it extends in to deeper layers. This affects the recombination wave in the SN simulation, and allows the photosphere to move more rapidly in the model creating a more rapidly evolving light curve. By more rapidly evolving we mean we see the same features, such as the peak, plateau, and tail, but they appear earlier. The abundances in figure 5 determine the opacities to be used in the explosion simulations. Opacities are taken from OPAL (Iglesias & Rogers, 1996). The extent of artificial mixing of $^{56}$Ni is not shown, but mixed out to the He core in all models. This is a secondary effect influencing the shape of the light curve. The initial radius and ejected mass are the primary parameters that affect the shape of the light curve. We do not include Rayleigh-Taylor instabilities, which would mix the compositional boundary layers. This effect is small and not expected to produce a significant change in the light curve.

The explosion simulation is performed in two steps as described by Young (2004). In an initial simulation, the hydrodynamics of each model is calculated and used to determine the deposition function for the gamma-ray emission in the double beta decays of $^{56}$Ni and $^{56}$Co. The second hydrodynamic simulation includes the energy from the gamma-rays as well as the generated shock wave. Table 4 shows the parameters of the starting models for the explosions. In the following sections we describe the outcome of the explosion parameters versus time for all models including for both the Schwarzschild and Ledoux convection criteria.

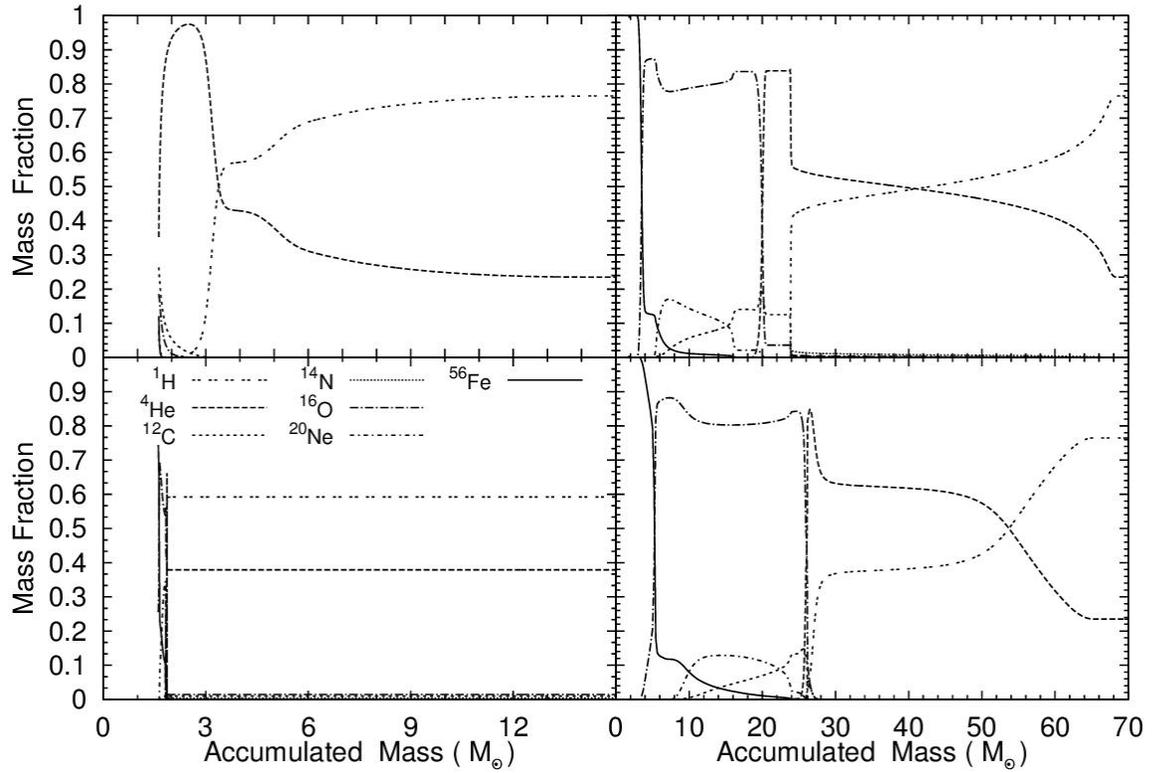

**Figure 5.** Stellar chemical composition profiles for 15 $M_\odot$ and 70 $M_\odot$ models, for both the Ledoux (top) and Schwarzschild (bottom) convention criteria just prior to core collapse.

**Table 4.** Parameters for the explosion models for both Ledoux (L) and Schwarzschild (S) models

| Model Mass/ $M_\odot$ | $R_L/R_\odot$ | $R_S/R_\odot$ | Explosion Energy L ($10^{51}$ ergs) | Explosion Energy S ($10^{51}$ ergs) | $^{56}$Ni mixing Mass/$M_\odot$ L | $^{56}$Ni mixing Mass/$M_\odot$ S |
|---|---|---|---|---|---|---|
| 15 | 8.74 | 603 | 1.42 | 1.40 | 2.52 | 1.86 |
| 20 | 9.57 | 341 | 1.26 | 1.25 | 7.00 | 3.15 |
| 25 | 12.12 | 567 | 1.50 | 1.50 | 5.00 | 4.70 |
| 30 | 12.06 | 1066 | 3.00 | 3.00 | 9.10 | 7.00 |
| 40 | 16.08 | 2048 | 2.34 | 2.32 | 13.65 | 11.59 |
| 50 | 20.51 | 520 | 3.80 | 3.80 | 17.78 | 16.80 |
| 60 | 22.59 | 1038 | 5.23 | 5.19 | 22.58 | 21.00 |
| 70 | 64.60 | 399 | 3.86 | 3.84 | 24.09 | 23.79 |
| 80 | 28.09 | 179 | 8.07 | 8.00 | 31.71 | 31.72 |

## 3.1 Stellar explosion results for Schwarzschild and Ledoux evolution models

The stellar explosion simulations change all physical parameters on a rapid time scale of days. All physical parameters including radius, density, velocity, pressure, temperature, and luminosity are followed throughout the simulation. The supernova luminosity is taken at the layer inside the star at which optical depth $\tau = 2/3$ in the expanding model and followed for 400 days.

In Figure 6 we show light curves for models with progenitor masses of 15 $M_\odot$ through 80 $M_\odot$, for both pre-explosion convection criteria. The most distinguishing features of the Schwarzschild light curves shown in the top panel are the bright peaks and smooth evolution to a long plateau, before dropping to a radioactive tail. The high initial peaks are due to the large initial model radii, whereas explosion energy, ejected mass and progenitor radius all contribute to the duration and brightness of the plateau. For the Schwarzschild models, there is no contribution from radioactive heating in the behavior of the plateau. Overall, there is no systematic relationship

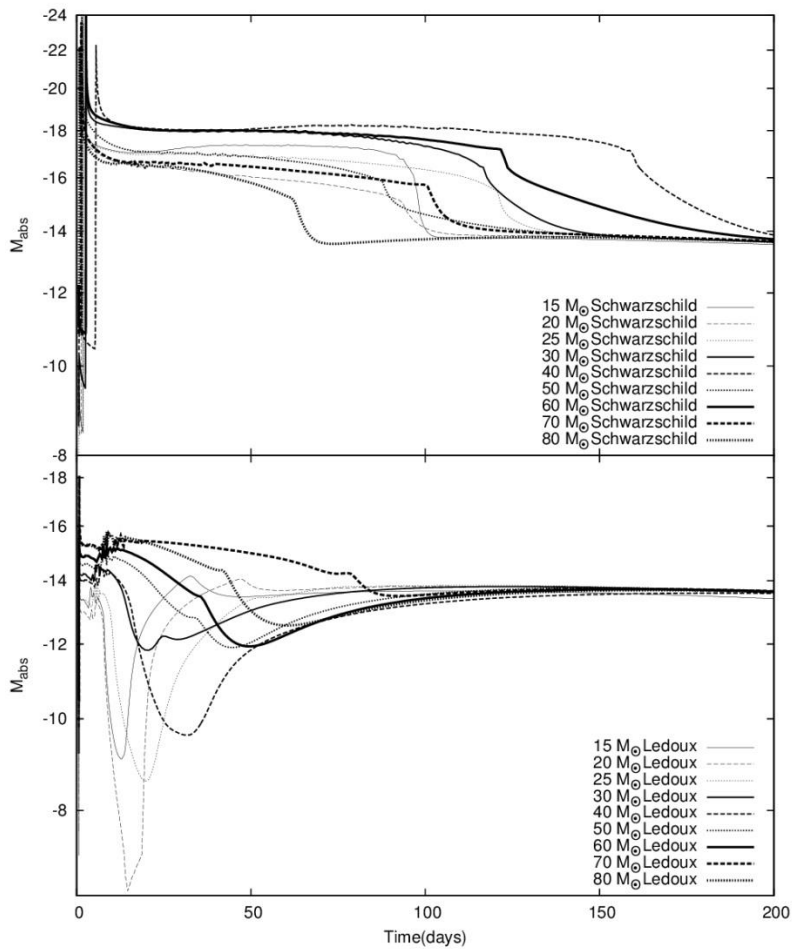

**Figure 6.** *Top:* Nine light curves for models using the Schwarzschild Criterion. *Bottom:* Nine light curves for models evolved with the Ledoux Criterion

between the model radius or mass and the shape of the light curve. This is because evolution models do not produce a correlation between pre-explosion radius and mass (see table 4). For example, evolution models from 20 – 40 $M_\odot$ show a monotonic increase in pre-explosion radius and the light curves show a corresponding increase in brightness during the whole 200 days. However; for our 40–80 $M_\odot$ evolution models, the radii do not follow any discernable trend, and the response in the light curves is an increase or decrease in brightness relative to radius. As expected, all light curves end at the same tail brightness because the same Ni mass was included in each model. Because the masses of all models are large, we expect all gamma-rays to be absorbed and contribute to the luminosity.

In the bottom panel of figure 6, we show light curves for the same mass range but for models evolved using the Ledoux criteria. The initial peaks in luminosity are very dim, with a shelf-like or flat appearance. The shapes of these light curves are primarily influenced by the smaller progenitor radii compared to the Schwarzschild models. Another striking difference in the light curve shapes for Ledoux models is the deep decline in luminosity seen between 20-70 days. This is again due to the small initial radii; most of the energy in the explosion goes into the gas expansion or *PdV* work. Interestingly, all but the 70 $M_\odot$ model are smaller than SN1987A, which had a similar light curve. The consequence of such small radii is that the majority of the light curve is powered by Ni decay, in contrast with the Schwarzschild models. The increase in the luminosity following the deep decline is caused by radioactive heating from $^{56}$Ni. The duration of the decline is determined by the time it takes for the recombination wave to reach the heated material. In most models, however, the $^{56}$Ni is mixed through the helium core and to the center of the star and so the heating is not as evident as a peak. But for these, the luminosity increase merges with the radioactive tail. This is true for the lower mass models, but in the higher mass models it is necessary to increase the initial explosion energy in order to achieve an explosion. For that reason some of the light curves for the larger radii models have plateau-like phase. The $^{56}$Ni influence can be seen later as a small bump at the end of the plateau for the larger masses. Again, all light curves converge to the same spontaneous radioactive tail that is determined by the input mass of $^{56}$Ni, 0.07 $M_\odot$. Although observed masses of $^{56}$Ni have been found to be above and below 0.07 $M_\odot$, this is the value determined for SN1987A a well-studied SN, and this mass is generally described as the canonical value (see for example, Sollerman 2002).

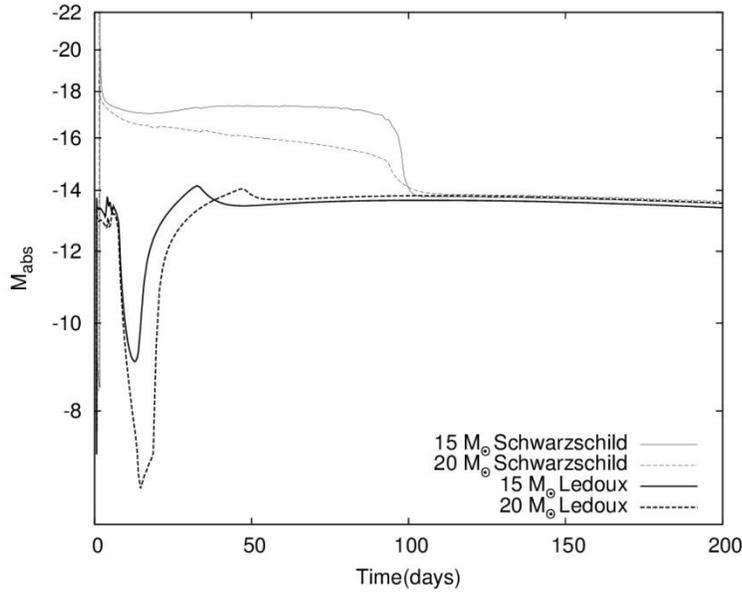

**Figure 7.** Light curves for both Schwarzschild and Ledoux evolution models of mass 15 and 20 $M_\odot$.

Figures 7-10 show the comparisons of light curves for models evolved with the two different convection criteria. Comparing light curves based on convection criteria for the same mass models shows that in general the Schwarzschild models result in a brighter light curve by about 4 magnitudes. Figure 7 shows the light curves for the 15 and 20 $M_\odot$ Schwarzschild and Ledoux models (S and L models hereafter). The pre-explosion S and L models have radii of 604 and 8.59 $R_\odot$ respectively. As expected, the radii have a dramatic effect on the light curve because most of the explosion energy in the Ledoux model is expanding the mass, decreasing the temperature, and produces a low luminosity event. This is even more extreme when compared to the SN1987A event which had a relatively large 43 $R_\odot$ progenitor radius in comparison. The S model is brighter and looks more like a typical supernova light curve. In each case the comparison to the 15 $M_\odot$ models give predictable results. The light curve for the 15 $M_\odot$ S model is brighter than the 20 $M_\odot$ S model, which can be explained by it having almost double the progenitor radius: 604 $R_\odot$ compared with 341 $R_\odot$. The L models do not have as large a difference in the progenitor radius. In fact their radii are very similar and the change in the light curve can only be explained by the difference in the ejected mass. The greater the mass ejected, the slower the light curve evolves and the features become more pronounced. To this point, the initial dip in the light curve is deeper and the secondary peak, caused by the reheating of the matter from Ni decay, is later.

Figure 8 shows light curves for the 25 and 30 $M_\odot$ S and L models. An increase in explosion energy for the 30 $M_\odot$ models causes significant differences in the appearance of the light curves. This is inevitable since the gravitational potential increases with increasing progenitor mass.

Thus, it is necessary to double the explosion energy to $3\cdot10^{51}$ ergs for both the 30 $M_\odot$ S and L models as compared with $1.5\cdot10^{51}$ ergs for the 25 $M_\odot$ models. This explains the brighter peak magnitude of the light curves and why they produce a faster light curve, for which features such as the secondary peak occur sooner. Another consequence is that the Ni heating appears earlier for the 30 $M_\odot$ L model due to higher velocity material, and thus the dip is not as deep as in the 25 $M_\odot$ L model. The combination of an increase in explosion energy and a larger progenitor radius for the 30 $M_\odot$ S model compared with the L models and the 25 $M_\odot$ S model makes it much brighter, constantly two magnitudes brighter than the 25 $M_\odot$ S model and almost 6 magnitudes brighter than both L models. The large difference in magnitude for most of the light curve out to 200 days between the S and L 30 $M_\odot$ models is due to the difference in radii: 1066 $R_\odot$ and 12 $R_\odot$.

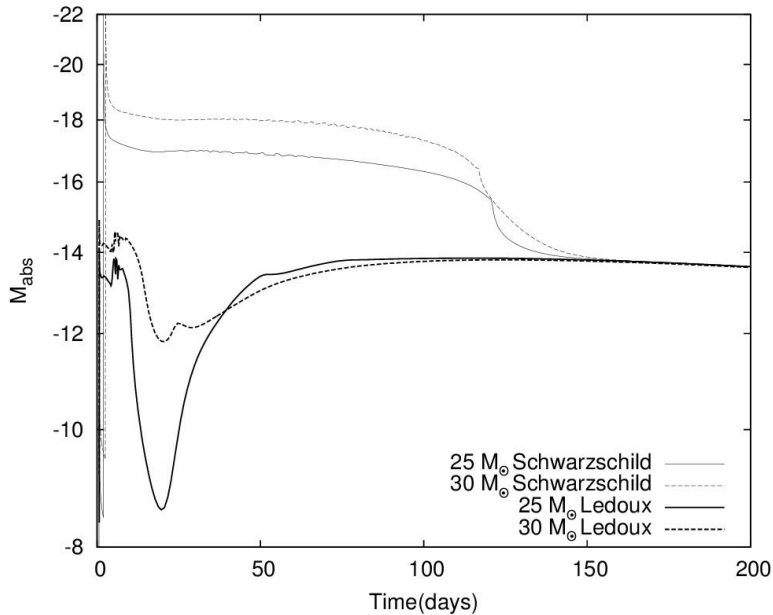

**Figure 8** Light curves for both Schwarzschild and Ledoux evolution models of mass 25 and 30 $M_\odot$.

Figure 9 shows light curves for 40 and 50 $M_\odot$ models. The model with the largest radius (2000 $R_\odot$) is the 40 $M_\odot$ S model, which is twice that of the 30 $M_\odot$ S model and four times that of the 50 $M_\odot$ S model. The 40 $M_\odot$ S model is our brightest supernova light curve model of all models, including those that are more massive. It also has the longest plateau duration, reaching out to roughly 160 days. This is true even though we use a lower explosion energy compared to that used for both the 30 and 50 $M_\odot$ models, which use $3.0\cdot10^{51}$ ergs and $3.8\cdot10^{51}$ ergs respectively, compared to $2.38\cdot10^{51}$ ergs for the 40 $M_\odot$ S model. Also shown in figure 9 are light curves for the 40 and 50 $M_\odot$ L models. The 50 $M_\odot$ L model has only a slightly larger progenitor radius than the 40 $M_\odot$ L model, but the light curve for the 50 $M_\odot$ L model has a noticeably brighter peak, which it is surprising given the additional ejected mass should make the initial light curve fainter. This is most likely due to the difference in explosion energy of the models: $2.34 \cdot 10^{51}$ ergs for the 40 $M_\odot$ L model compared to $3.8 \cdot 10^{51}$ ergs for the 50 $M_\odot$ L model.

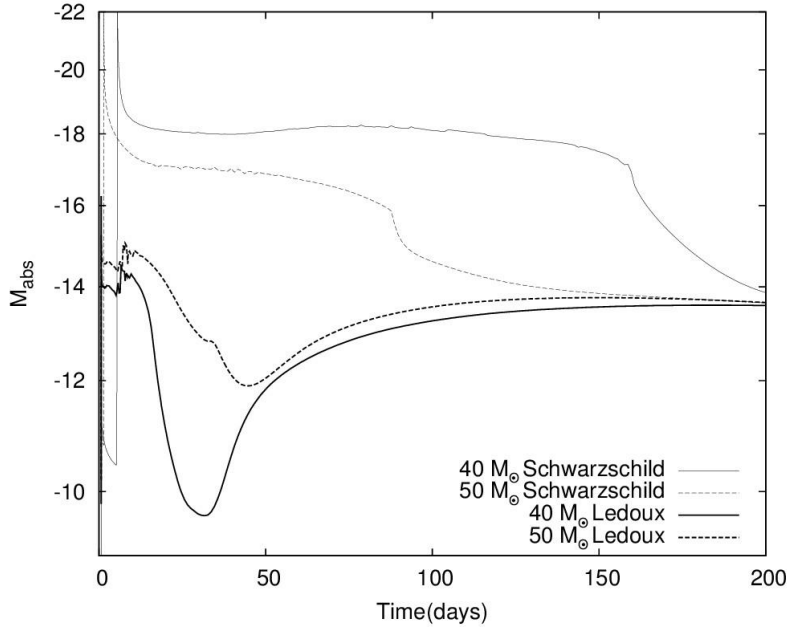

**Figure 9** Light curves for both Schwarzschild and Ledoux evolution models of mass 40 and 50 $M_\odot$.

Figure 10 shows light curves for both S and L models for 60, 70 and 80 $M_\odot$. We vary the energies widely in order to achieve an explosion for each model. The 60 $M_\odot$ models need roughly $5 \cdot 10^{51}$ ergs, and the 70 and 80 $M_\odot$ models require $4 \cdot 10^{51}$ and $8 \cdot 10^{51}$ ergs, respectively. We use the minimum energy to produce an explosion, which ultimately depends on the structure of the individual cores of the models. The 60 $M_\odot$ S model has the largest progenitor radius in this sequence and is the brightest, while the 70 and 80 $M_\odot$ models have smaller radii and are fainter. Likewise, the brightest light curve for the L model sequence is for the 70 $M_\odot$ L model which also has the largest progenitor radius of 54 $R_\odot$. The next brightest should be the 60 $M_\odot$ L model which has a radius comparable to the 80 $M_\odot$ L model and results in less ejected mass, but because the explosion energy for the 80 $M_\odot$ L model is double, it is brighter.

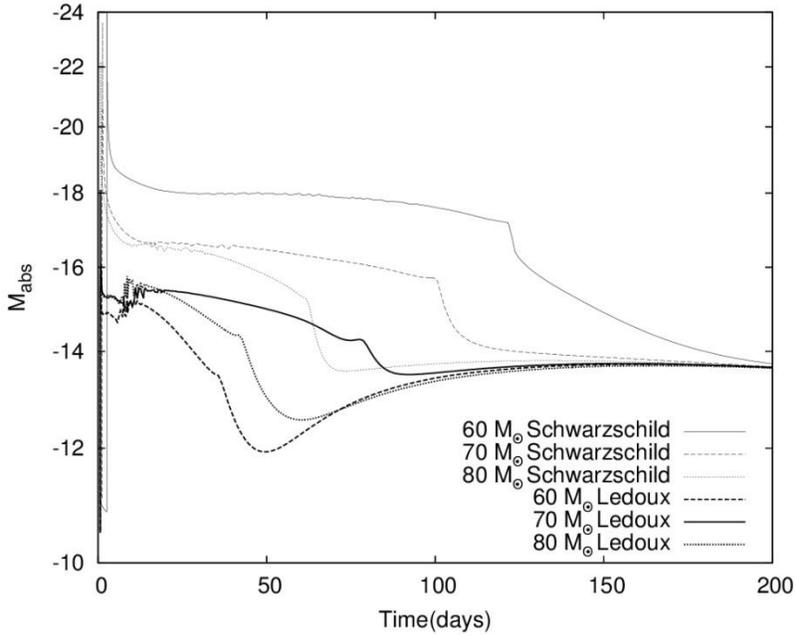

**Figure 10** Light curve comparison for Schwarzschild and Ledoux models for 60-80 M$_\odot$.

## 3.2 Supernovae at high redshift and the k-correction for Schwarzschild and Ledoux stellar models

We simulate the appearance of our supernovae light curves at a very early time in the universe by using a high redshift and applying a k-correction as described by Kim et al., 1996. Using the k-correction given by,

$$k = -2.5\log_{10}\int fB_\lambda\left(\frac{\lambda}{1+z}\right)d\lambda,$$

we calculate the adjusted apparent magnitude,

$$m = 5\log_{10}(d) - 5 - 5\log_{10}(1+z) + k.$$

Figure 11 shows the brightest supernova light curve, our 40 M$_\odot$ S model, at redshifts of $z = 0, 5, 10$, and 15. In order to compare to what would be realistic observations, we integrate the Planck function derived from the simulations over the bands z, j, h, k, l, and m. We use these longer wavelength bands because there is little flux in bands B, V, R, I at high redshift and so these are not shown. The top left panel of figure 11 shows the Johnson band filter light curves for z, j, h, k, l, m bands for the 40 M$_\odot$ S model as it is would look in the rest frame of the supernova. Shown in the remaining three panels, the redshifted light curves for $z = 5, 10$ and 15 are progressively attenuated and distorted. During the peak maximum brightness of the light curves, all bands are bright with a slow dispersion with higher redshift favoring more flux in longer wavelength bands. The k band remains the brightest during the plateau phase for $z = 5$ and 10 while at $z = 15$ the m-band is the brightest.

For comparison, light curves for the Ledoux convection models are presented in figure 12 for $z = 0$ and 15. The most dramatic difference is that the Ledoux light curves are fainter and still have the dim peak, subsequent decline and final rise to the tail, as seen earlier in the bolometric light curve. At our highest redshift, $z = 15$, the l and m bands have moved to be the brightest bands. A difference between the S and L model light curves is that the L model light curves vary less over time for each filter band, whereas for the S model light curves one filter band (the z band) is bright during the peak but falls fainter on the plateaus than all other filter bands. This has consequences for potential observations through the duration of the light curve peak and plateau. It can predict which filters one would need to observe real primordial SN. The physical mechanism that causes this difference is that the S models have a higher temperature early during the peak and then cool during the plateau. The L models are already cool to start due to the initial expansion experienced by those models.

In Figure 13 we show light curves for all S (bold lines) and L (standard lines) models in z' and 2mass Ks filter bands along with the James Web Space Telescope's (JWST) limiting magnitude in the 1–5 μm range at m = 33 redshifted to $z = 5$. Most Ledoux models (15–70 M$_\odot$) fall below the liming magnitude and would be difficult to observe even at $z = 5$. However the 80 M$_\odot$ L model in the z' and Ks filter band are above the detection limit for 5 days. The Schwarzschild models show a larger observing window ranging from 12 to 25 days for the z'-band. Interestingly for the 30, 40, and 60 M$_\odot$ S models, the light curves stay at the detection limit for much of the plateau in the Ks-band.

Figure 14 shows light curves for all S (bold lines) and L (standard lines) models in z' and 2mass Ks filter bands with the James Web Space Telescope (JWST) limiting magnitude in 1–5 um range at m=33 redshifted to $z = 15$. All Ledoux models fall below the limiting magnitude and would be extremely difficult to observe at $z = 15$ in both filters. At this high redshift, The S model light curves show a very narrow observing window which never gets out 1 day, even for the brightest models. However, longer observing integration times would most favor detection for the 40 M$_\odot$ and 60 M$_\odot$ mass models in both z' band and Ks band.

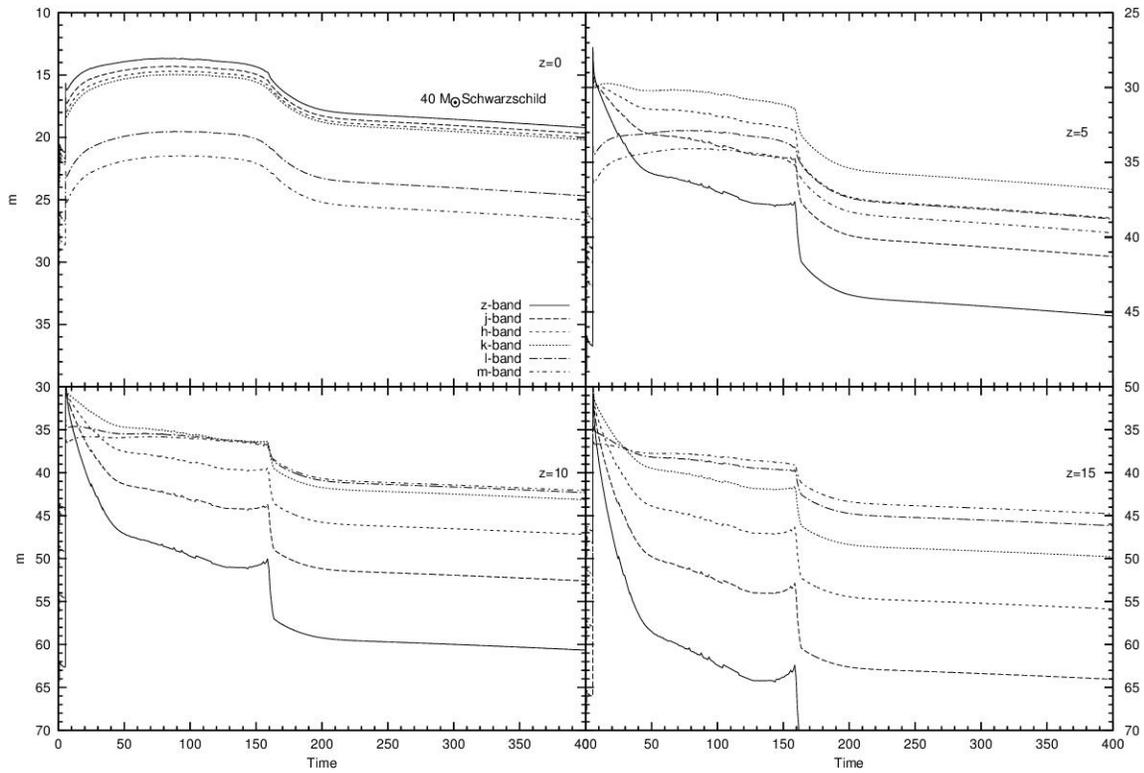

**Figure 11** Light curves for the 40 M$_\odot$ S model in z, j, h, k, l, and m bands. Top left panel has redshift $z = 0$, top right has $z = 5$, bottom left has $z =10$, and bottom right has $z = 15$.

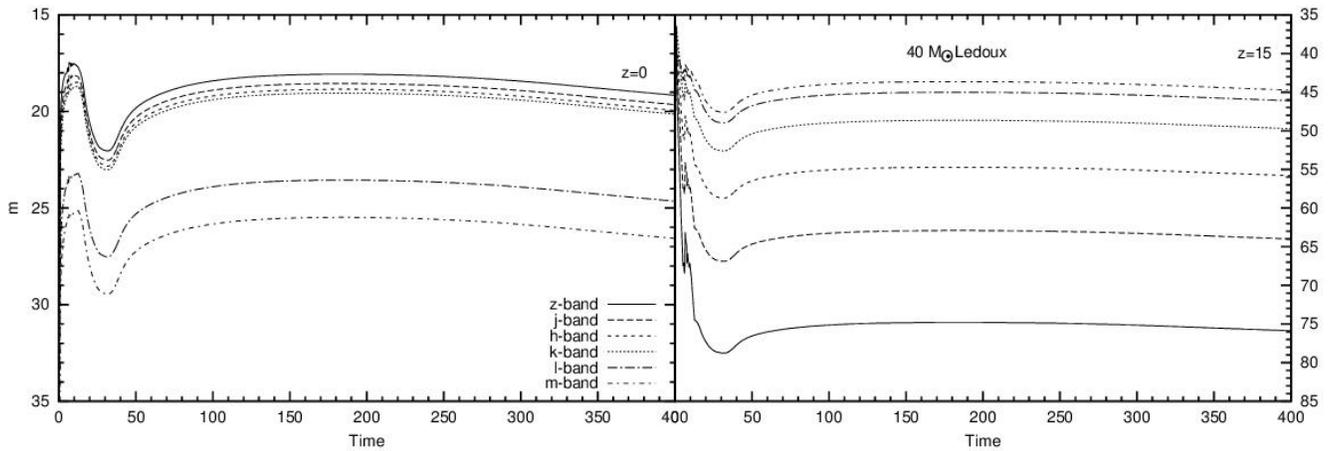

**Figure 12** Light curves for the 40 M$_\odot$ L model in z, j, h, k, l, and m bands. The left panel is for redshift z = 0 and right is for redshift $z = 15$.

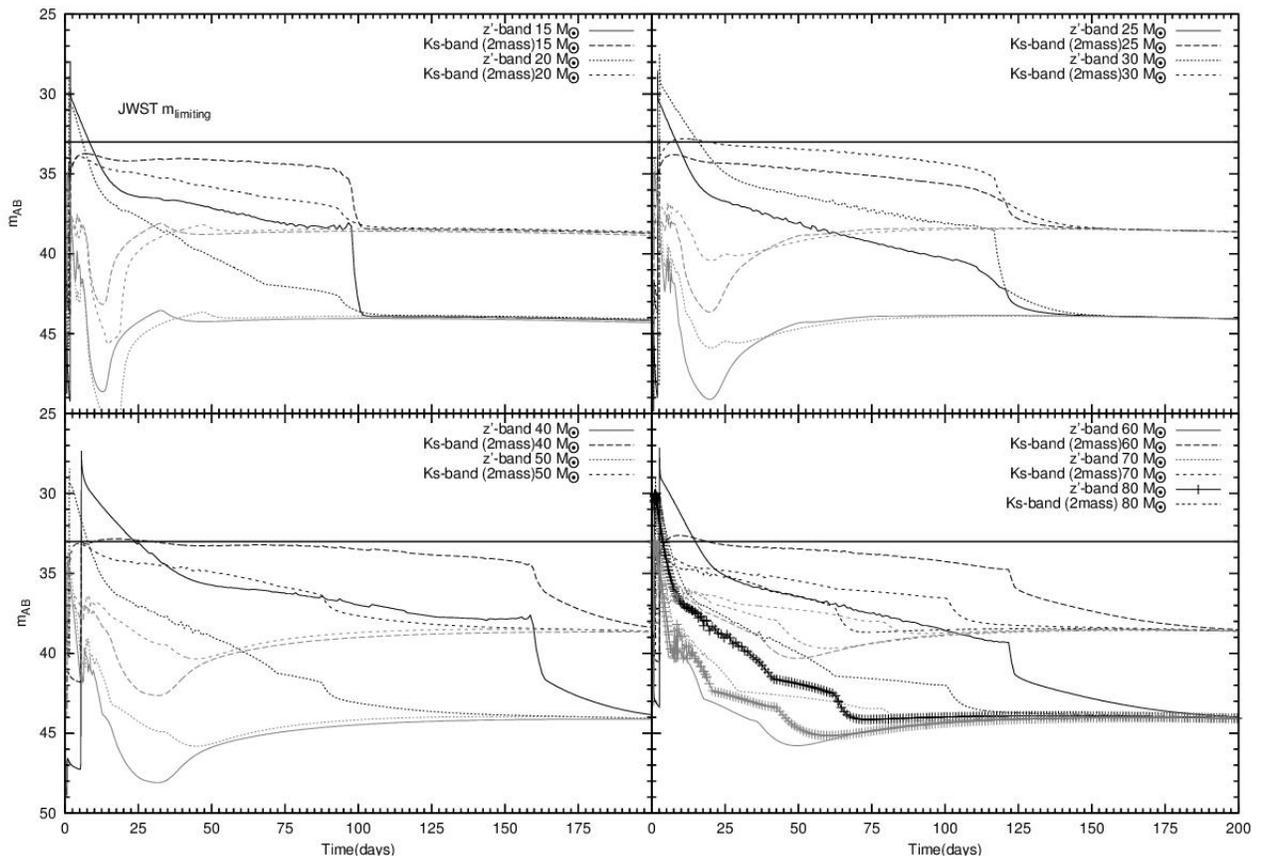

**Figure 13** Light curves for S (bold lines) and L (standard lines) models with redshift $z = 5$. The top left panel shows 15 and 20 $M_\odot$ S and L models in Johnson z and 2mass Ks filter bands. The top right panel shows 25 and 30 $M_\odot$ S and L models in Johnson z and 2mass Ks filter bands. The bottom left panel shows 40 and 50 $M_\odot$ S and L models in Johnson z and 2mass Ks filter bands. The bottom right panel shows 60, 70 and 80 $M_\odot$ L and S models in Johnson z and 2mass Ks filter bands.

In general the light curves show a more pronounced peak at high z than seen in the uncorrected light curves. This feature in the light curve is due to the $(1 + z)$ term in the k-correction wavelength shift of the models. The energy is shifted to longer wavelengths. The best possibility of detection would occur at the peak of the light curve above the limiting magnitude. However, with longer integration times it may be possible to detect the SN during the longer duration plateau phase. The 40 $M_\odot$ S model in the z' filter band is brightest and has the longest in duration above the limiting magnitude. If the limiting magnitude is pushed to $m = 35$, the model stays above for longer duration in the 2mass Ks filter band.

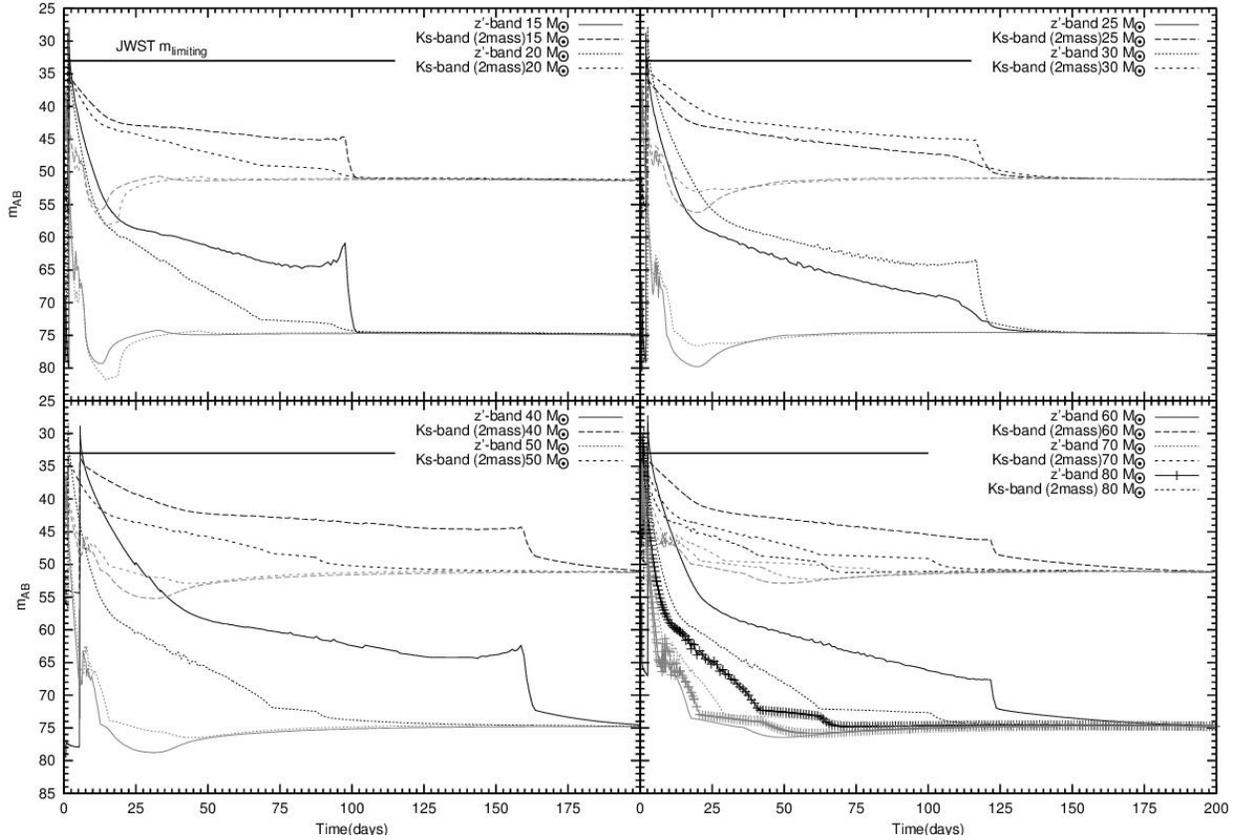

**Figure 14** Light curves for S (bold lines) and L (standard lines) models with redshift $z = 15$. The top left panel shows 15 and 20 $M_\odot$ S and L models in Johnson z and 2mass Ks filter bands. The top right panel shows 25 and 30 $M_\odot$ S and L models in Johnson z and 2mass Ks filter bands. The bottom left panel shows 40 and 50 $M_\odot$ S and L models in Johnson z and 2mass Ks filter bands. The bottom right panel shows 60, 70 and 80 $M_\odot$ S and L models in Johnson z and 2mass Ks filter bands.

## 4 SUMMARY AND DISCUSSION

To study how the choice of treatment of convection in modelling pre-supernovae evolution influences the predicted light curves of population III core-collapse supernovae, we have evolved models of population III stars in the mass range 15 $M_\odot$ to 100 $M_{\odot\epsilon}$ from the pre-main sequence to the initial stages of photo-disintegration of the iron core for two convection models. We find that using the Schwarzschild convection criterion leads to model stars that are significantly larger and cooler at core-collapse than models evolved using the Ledoux convection criterion. This difference is primarily due to the presence of a molecular weight gradient outside the helium burning core. In models using the Schwarzschild criterion, convection moves inwards after the end of core helium burning, allowing mixing of hydrogen into hot carbon rich layers. The ensuing phase of increased hydrogen burning causes expansion of the envelope. In models using the Ledoux criterion, the molecular weight gradient prevents inward movement of the convection zone and mixing of hydrogen into carbon-rich layers. We also compare our models to those of Heger & Woosley (2010) who used the Ledoux criterion but also included semi-convective mixing and some convective overshoot. All our Ledoux convection models at core collapse are smaller and hotter than the corresponding models of HW, and, in general, our Schwarzschild convection models at core collapse are larger than the corresponding models of HW. Thus our models to a major extent bracket the range of behaviours that result from different prescriptions for convection. A possible exception is inclusion of large amounts of convective overshoot.

The convective treatment in our evolution models has a significant effect on their final radii, which in turn determines the shape and largely determines the brightness of our simulated supernova light curves. Overall, the Schwarzschild models produce pre-explosion radii larger than those using the Ledoux criterion. In fact, the 40 $M_\odot$ Schwarzschild model produces the largest pre-explosion radius in the entire mass range of 15 – 80 $M_\odot$. This is vitally important in the context of simulated SN light curves because light curves depend critically on the pre-explosion radius. We have also compared light curves that include the applied k-corrections due to high redshift. We find here that potential detection is more favorable for the Schwarzschild convection criterion and surprisingly includes intermediate mass stars. The combination of larger mass and particularly large radius for our 40 $M_\odot$ S model produces a plateau phase in the light curve with the longest duration. The red-shifted supernovae light curves produced from Schwarzschild convection models with k-corrections are brightest in the k-band early, but after a short time switch to being brightest in the m-band. The switch is dramatic in that the other bands are never the brightest at any time.

From our light curve results for 15 – 30 $M_\odot$ progenitor models, we predict that it would be extremely difficult to observe SN of these progenitor masses since their luminosities are faint for the entire duration of time, except for a few early days, and only for the light curves at redshift $z$ = 5 – 10. The predicted detectability of SN for 50 $M_\odot$ through 80 $M_\odot$ progenitors diminishes, with the 80 $M_\odot$ model light curve dropping below detection limits in less than a day. At redshifts $z$ = 10 – 15, the most promising filter bands for detection are the k and m-bands for observing a supernova during the longer plateau region rather than during the peak. This is also true for the JWST in the z' and Ks filter bands. We show that most models produce faint light curves and that with k-corrections, these objects would become difficult to observe. Thus, it may not be surprising that no primordial SNe have yet been observed. Based on our results, we conclude that in order for the JWST, for example, to see a type II core collapse SN, it would be due to a progenitor star that is 40 $M_\odot$ or possibly 60 $M_\odot$ (of the S variety) rather than lower or higher mass progenitors, although this depends on which convection criterion is closer to 'right' one. We show that all Ledoux model light curves are fainter by 4 – 6 magnitudes compared to the Schwarzschild model light curves, and fall well below the detection limits of JWST. Thus, the lower limit for mass of the first stars may be higher if the Ledoux Criterion is closer to the 'right' criterion. This prediction could be used as an indirect test to determine a limit for the initial mass function (IMF) for the first stars: 1.) If the JWST does not see any type II core collapse SN in distant galaxies, it may suggest that the first star IMF is lower than previously expected, though how much lower would be speculative, or it may mean that their radii are smaller; that is, that they evolve with something more closely resembling the Ledoux criterion. 2.) We predict that if only few SNe are observed, then the initial mass for the first stars can either be in the range 35 – 45 $M_\odot$ or bimodal around 40 $M_\odot$ and 60 $M_\odot$. 3.) If JWST detects copious core collapse SNe for long integration times in the Ks filter bands, we predict a wider intermediate IMF range of 30 $M_\odot$ – 70 $M_\odot$, as those model light curves are equally bright during the plateau and thus would be more likely to observe.

# REFERENCES

Böhm-Vitense E. 1958, ZA, 46, 108
Bromm V., Larson R.B., 2004, Annual Review of Astronomy and Astrophysics, 42, 79
Bromm V., Loeb A., 2003, Nature, 425, 812
Bouwens R. J. et al., 2011, ApJL, 736, L28
Chieffi A., Dominguez I., Hoflich P., Limongi M. Straniero O., 2003, MNRAS, 345, 111
Coe D. et al., 2013 ApJ, 762, 32
Dunlop J. S., 2012, arXiv:1205.1543
Eckström S., Coc P.D., Meynet G., Olive K.A., Uzan J.-P., Vangioni E., 2010, A&A, 514, A62
Eggleton P.P. 1972, MNRAS, 156, 361
Ezer D., Cameron A. G. W., 1971, Ap&SS, 14, 399
Eryurt-Ezer D., Kiziloglu N., 1985, A&SS, 117, 95
Figer D. L., Rauscher, B. J., Regan, M. W., Morse, E., Balleza, J., Bergeron, L., Stockman., H.S., 2004, in Grycewicz, T. J., McCreight, C. R., eds., Proc. of SPIE Vol. 5167, Focal Plane Arrays for Space Telescopes , Bellingham, WA, p. 270
Glover S., 2005, Space Sci.Rev., 117, 445
Heger A., Woosley S. E., 2002, ApJ, 567, 532
Heger A., Woosley S. E., 2010, ApJ, 724, 341
Kato S., 1966, PASJ, 188, 374
Iglesias C.A., Rogers F.J., 1996, ApJ, 464, 943
Kim A., Goobar A., Perlmutter S., 1996, PASP, 108, 190
Langer N., Sugimoto D., & Fricke K.J., 1983, A&A, 126, 207
Lawlor T.M., MacDonald J., 2006, MNRAS, 371, 263
Lawlor T.M., Young T.R., Johnson T.A., MacDonald J., 2008, MNRAS, 384, 1533
Limongi M., Chieffi A., 2012, ApJS, 199, 38
Limongi M., Chieffi A., 2005, ASPC, 342, 122
Maeder A., 1997 A&A, 321, 134
Marigo P., Girardi L., Chiosi C., Wood P. R., 2001, A&A, 371, 152
Mihalas D. 1978, Stellar Atmospheres (2nd ed.; San Francisco, CA: Freeman)
Oesch P. A. et al., 2010, ApJL, 709, 21

Palla F., Stahler W., Salpeter E.E., 1983, ApJ, 271, 632
Plank Collaboration, 2014, A&A, 571, 30
Reimers D., 1975, Mem. R. Soc. Sci. Liege, 8, 369
Rydberg C., Zackrisson E., Lundqvist P., Scott P., 2013, MNRAS, 429, 3658
Schneider R., Ferrara A., Salvaterra R., Omukai K., Bromm V., 2003, Nature, 422, 869
Siess L., Livio M., Lattanzio J., 2002, ApJ, 570, 329
Sollerman J., 2002, New Astron Rev, 46, 493
Spruit H.C., 1992, A&A, 253, 131
Stacy A., Greif T.H., Bromm V., 2010, MNRAS, 403, 45
Sutherland P. G., Wheeler J. C., 1984, ApJ, 280, 282
Tanvir N.R. et al., 2009, Nature 461, 1254-1257
Turk M., Abel T., O'Shea B., 2009, Science, 325, 5940, 601
Weaver T.A., Zimmerman G.B., & Woosley S.E., 1978, ApJ, 225, 1021
Whalen D. J., Heger A., Chen K.-J., Even W., Fryer C. L., Stiavelli M., Xu H., Joggerst, C. C., 2012, arXiv:1211.1815 [astro-ph.CO]
Wright E. L., 2006, PASP, 118, 1711
Young T. R., 2004, ApJ, 617, 1233